\documentstyle[epsfig]{mn}
\font\japit = cmti10 at 10truept
\title
     [Reconstructing redshift surveys]
{\vglue-3.0truecm
\centerline{\japit Submitted to Monthly Notices}
\vglue 2.5truecm
\noindent
	The inverse redshift--space operator:
	reconstructing cosmological density and velocity fields
\author
     [A. N. Taylor \& H. Valentine]
     {Andy Taylor \& Helen Valentine\\
     Institute for Astronomy, 
     University of Edinburgh,
     Royal Observatory,
     Blackford Hill, 
     Edinburgh, 
     U.K.\\
        ant@roe.ac.uk,hemv@roe.ac.uk}}

\newcommand{\be}{\begin{equation}}
\newcommand{\ee}{\end{equation}}
\newcommand{\ba}{\begin{eqnarray}}
\newcommand{\ea}{\end{eqnarray}}

\newcommand{\q}{{\bmath q}}
\newcommand{\r}{{\bmath r}}
\newcommand{\rbh}{\hat{\bmath r}}
\newcommand{\rh}{\hat{r}}
\newcommand{\rhat}{\hat{r}}
\newcommand{\sbh}{\hat{\bmath s}}
\newcommand{\sh}{\hat{s}}

\newcommand{\0}{\bmath{0}}

\newcommand{\xib}{{\bmath \xi}}
\newcommand{\vb}{{\bmath v}}
\newcommand{\Y}{{\bmath Y}}
\newcommand{\G}{{\bmath G}}
\newcommand{\Wb}{{\bmath W}}
\newcommand{\Ylm}{Y_{\ell m}}
\newcommand{\Lb}{{\bmath L}}

\newcommand{\omegab}{{\bmath \omega}}

\newcommand{\hMpc}{h^{-1}{\rm Mpc}}

\newcommand{\de}{\partial}
\newcommand{\s}{{\bmath s}}
\newcommand{\Sb}{{\bmath S}}
\newcommand{\lmdk}{\sqrt{\frac{2}{\pi}} \sum_{\ell m} \int \! dk k^2 \,}
\newcommand{\rgl}{\rangle}
\newcommand{\lgl}{\langle}

\newcommand{\nn}{\nonumber \\}
\newcommand{\kms}{{\rm km}s^{-1}}

\def\bib{\parskip=0pt\par\noindent\hangindent\parindent
    \parskip =2ex plus .5ex minus .1ex}
\begin{document}

\maketitle

\begin{abstract}
	We present the linear inverse redshift space operator
which maps the galaxy density field derived from redshift surveys 
from redshift space to real space. Expressions are presented for 
observers in both the CMBR and Local Group rest frames. We show
how these results can be generalised to flux--limited galaxy redshift 
surveys. These results allow the straightforward reconstruction of 
real space density and velocity fields without resort to iterative or 
numerically intensive inverse methods.
As a corollary to the inversion of the density in the Local Group rest 
frame we present an expression for estimating the real space velocity 
dipole from redshift space, allowing one to estimate the Local Group 
dipole without full reconstruction of the redshift survey. We test these
results
on some simple models and find the reconstruction is very accurate.
A new spherical harmonic representation of the redshift distortion and 
its inverse is developed, which simplifies the reconstruction and
allows analytic calculation of the properties of the reconstructed 
redshift survey.
We use this representation to analyse the uncertainties in the reconstruction
of the density and velocity fields from redshift space, due to only a finite 
volume being available. Both sampling and shot-noise variance terms are derived 
and we discuss the limits of reconstruction  analysis.
We compare the reconstructed velocity field with the true velocity field and 
show that reconstruction in the Local Group rest frame is preferable,
since this eliminates the major source of uncertainty from the dipole mode.
These results can be used to transform redshift surveys to real space
and may be used as part of a full likelihood analysis to extract cosmological
parameters.

\end{abstract}


 
\section{Introduction}

Galaxy redshift surveys are distorted by the intrinsic 
motions of galaxies along the line of sight. This distortion 
leads to a number of effects. On large, linear scales 
the radial displacement of galaxy
positions is enhanced, leading to compressed structures along
the line of sight. On smaller scales, near turnaround, the 
structure will appear completely collapsed along the line of
sight, and thereafter will begin to invert itself. Finally on very
small scales, after virialisation, the random motions of galaxies
in clusters will stretch out the structure in redshift space.

These distortions can complicate measurements of the statistical 
properties of the large scale galaxy distribution, and can lead 
of confusion in the interpretation of redshift maps. However, the
distortions also contain valuable information on the structure
of cosmological velocity fields. Furthermore, in gravitational instability
theory the parameter controlling the degree of distortion is a function of
the mean density of the Universe (see the excellent review by Hamilton, 1997,  
on linear redshift distortions).

One of main problems of cosmography is how to map the large
scale structure seen in galaxy catalogues from redshift space to real 
space. In recent years there has been much interest in reconstructing 
redshift surveys, with the advent of the IRAS all-sky surveys, QDOT and
1.2Jy,
and in particular the new Point Source Catalogue redshift survey
(hereafter PSCz; see Saunders et al 1996). While it is straightforward
to write down the transformation from real space to redshift space
galaxy distributions, the inverse relation has proven more
elusive.

The standard approach to reconstructing real space and the velocity fields
 has been to apply some iterative method,
to solve for the  density field (Kaiser et al 1991, 
Taylor \& Rowan-Robinson 1993, Branchini et al 1998) or reconstruct galaxy 
positions, again iteratively (eg, Yahil et al 1991). Peebles (1989)
showed that the trajectories of galaxies could be solved numerically 
by a Least Action
principle and Shaya et al (1993) applied this to large scale structure.
Recently Croft \& Gaztanaga
(1997) have shown that applying the Zel'dovich approximation to 
a Least Action principle leads to a combinatorics problem of 
finding the least path between initial and final conditions. 
Valentine et al (1999) have generalised this to selection
function weighted galaxy surveys and applied the method to the 
Point Source Redshift Catalogue. Again, the process is iterative
and may be time consuming. Going to a higher order reconstruction 
appears to be a slow and painful process. 

 An alternative method has been to expand the density field in spherical
harmonics and solve for each multipole, either from a  numerically intensive
inversion method (Fisher et al 1995, Zaroubi et al 1995, 
Webster et al 1997), or by numerically 
solving the redshift space Poisson equation to find the velocity potential
field (Nusser \& Davis 1994). The former method has the advantage of 
calculating the uncertainties in the reconstruction, but assumes a Gaussian 
distributed density field and requires regularization in the form of
a Wiener filter. The latter has the advantage of being extendable into
the Zel'dovich regime, but again must be solved numerically if
flux--limited.

Finally, Tegmark \& Bromley (1995) proposed 
a Green's function solution to the linear inversion problem, allowing one 
to regain the real space density and velocity fields from a straightforward 
convolution. However their method was only valid for volume limited galaxy 
distributions.

With the advent of the PSCz and forthcoming completion of the AAT 2dF 
(Maddox et al 1998) and Sloan Digital Sky Survey (Szalay et al 1998) 
an accurate, but straightforward
method for removing the distortion in flux--limited redshift surveys is required.
Reconstruction can be seen as required for comparison of velocity 
field data with the inferred flow in redshift surveys, or a general
method for correcting the distortion before subsequent analysis.

Since redshifts are taken in the rest frame of the
observer, the inversion method should deal with the local dipole motion
of the observer. The dipole itself is of interest, since a
reconstruction of the real space dipole can be compared with the 
observed CMB dipole and again put constraints on the mean density of
the Universe (Rowan-Robinson et al 1990, Kaiser \& Lahav 1988, 
Rowan-Robinson et al 1999, Schmoldt et al 1999). Since the dipole is 
nonlocal, the 
whole volume of the redshift survey usually has to be reconstructed in order to 
estimate just the dipole.

In this paper we address the problem of linear reconstruction
of galaxy redshift surveys, and put it on a firm theoretical foundation.
In Section \ref{mapping} we derive the inverse
spherical redshift distortion operator by considering the displacement
fields of galaxies in real and redshift space. This operator
allows one to transform directly between the linear redshift 
density field and the linear real space density field, without 
iteration. This allows a quick and simple, yet accurate, method for removing
the linear distortion seen in redshift maps. We
derive versions of this inverse operator for volume and flux--limited
redshift surveys and for observers in the Cosmic Microwave Background
(CMB) and Local Group rest frames.
As a corollary of the Local Group inverse redshift operator we 
derive an expression for estimating the real space observer's dipole
motion directly from a redshift survey, without a full inversion. This 
removes the effects of the
redshift distortion and corrects for the well--known ``Rocket effect''
(Kaiser \& Lahav 1988).

We develop a spherical harmonic representation of the distortion
operator and its inverse in Section \ref{sphharm}. This allows both practical and 
convenient application of the operators to redshift surveys, and
simplifies the calculation of the properties of the reconstructed 
density and velocity fields. We demonstrate the method by reconstructing
both linear density fields and the real space dipole from a 
simple model of a redshift survey.

In Section \ref{prop} we use our spherical harmonic formalism to estimate the 
uncertainties in the reconstruction of the density and velocity fields
with only a finite survey volume. We derive the
statistical properties of the reconstructed velocity field, taking
into account cosmic variance and shot noise. This allows us to
place limits on the accuracy to which one can do such reconstructions.

We summarise our results and present our conclusions 
in Section \ref{conclusions}.

\section{Mapping from redshift--space to real--space}
\label{mapping}

The problem of finding an inverse redshift--space operator depends
on the specifics of the redshift survey to which it is to be applied.
The redshift survey may be volume-- or flux--limited. In the case of
the latter, the mean observed density of galaxies is a function of 
distance from the observer and requires an extra correction. The 
redshifts of the galaxies in a survey are taken in the observer's
rest frame, and so include the motion of the Earth with respect 
to the CMB. Traditionally this has been transformed to either
the Local Group rest frame -- where the Local Group represents the 
largest gravitationally bound structure to which we belong, and 
whose size is close to the linear regime -- or the CMB rest frame.
In the rest of this paper we shall refer to the ``Local Group'' rest
frame in rather a loose way, referring to a hypothetical observer
moving in the local linear flow field with respect to 
the CMB rest frame. 

In the next few sections we find solutions to the inverse distortion in
the CMB frame for volume--limited (Section \ref{vollimited}) and flux--limited
(Section \ref{fluxlim}) surveys, and in the Local Group frame 
(Section \ref{LGframe}).
In addition the effects of the ``Local Group'' motion on redshift space and
its inverse operator lead us to an expression for the calculating
the real space dipole from a redshift survey (Section \ref{LGframe}).

We begin by considering the linear displacement of galaxies and
derive the inverse redshift operator for a volume limited redshift 
survey.

\subsection{Volume limited redshift inversion}
\label{vollimited}

The forward problem, of mapping a density field from real 
space to redshift space, can be solved using the transformation\footnote{
Throughout we shall use velocity units for distances. This is equivalent to
setting the Hubble parameter $H_0=1$ in the formulae.}
\be
	\s=\r+ (\rbh.\vb) \rbh
\ee
where $\s$ and $\r$ are the comoving redshift and real-space coordinates,
and $\vb$ is the peculiar velocity field. In linear theory the velocity
field is calculated from the density field by (Peebles 1980)
\be
\vb(\r) = \frac{f(\Omega_m)}{4 \pi} 
	\int \! d^3\! r' \, \delta(\r) \frac{\r'-\r}{|\r'-\r|^3}
\label{eqvel}
\ee
where
$f(\Omega_m) \simeq \Omega_m^{0.6}$ is the growth rate of the density field
and $\Omega_m$ is the mass density parameter of the Universe.

It is useful to introduce a displacement vector field, $\bxi$, defined by
\be
	\frac{d\bxi}{dt} \equiv \vb 
\ee
which denotes the displacement in real space from some 
initial, or Lagrangian, position
to the present day. In linear theory the velocity is proportional to
the displacement,
\be
	\vb = f(\Omega_m) \xib.
\ee
To allow for the possibility of linear biasing of the galaxy density
field we introduce the linear bias factor, $b$,  defined by 
$\delta_g=b \delta_m$. If velocities are estimated from the galaxy 
distribution, rather than the true density, the constant of
proportionality between velocities and displacements is $\beta$
where $\beta\equiv\Omega_m^{0.6}/b$.

The real space and redshift space coordinates, $\r$ and $\s$ respectively,
can be related to a set of initial or Lagrangian coordinates, $\q$, by
\be
	\r = \q + \xib,
\ee
and
\be
	\s = \q + \xib^s.
\ee
The redshift space displacement field, $\xib^s$, is related to the 
real space displacement, $\xib$, by (Taylor \& Hamilton 1996)
\be
	\xi^s_i = {\cal P}_{ij}\xi_j
\label{real2red}
\ee
where 
\be
	{\cal P}_{ij} \equiv \delta^K_{ij} + \beta \rh_i \rh_j
\ee
is the redshift space projection tensor, and $\delta^K_{ij}$ is the 
Kronecker tensor. From continuity we can
find the linear redshift density by taking the divergence of
the redshift displacement field:
\be
	\delta^s = - \nabla. \xib^s.
\label{divred}
\ee 
Substituting equation (\ref{real2red}) into equation (\ref{divred}) yields
\ba
	\delta^s &=& -(\nabla.\xib + \beta 
		(\rh_i \rh_j \nabla_i \xi_j + 2r^{-1} \rh_i \xi_i)) \nn
		&=& [1+\beta (\de^2_r+2r^{-1} \de_r) \nabla^{-2}] \delta
\label{redcont}
\ea
where we have used the real space continuity equation
\be
	\delta = - \nabla.\xib,
\ee
and $\nabla^{-2}$ is the inverse Laplacian. To linear order
spatial derivatives in redshift space are equal to derivatives in 
real space, and the coordinates of fields are the same
in both real and redshift space, i.e. $\delta(\s)=\delta(\r)$.

Equation (\ref{redcont}) can be written much more
succinctly if we introduce the linear, spherical redshift space 
operator (Hamilton 1997)
\be
	\Sb = 1+\beta (\de^2_r + 2r^{-1} \de_r) \nabla^{-2},
\ee
and write the transformation from real to redshift density 
fields as 
\be
	\delta^s = \Sb \,\delta.
\ee

Having formulated the redshift distortion in the language of 
displacements
the inverse operator, $\Sb^{-1}$, is now straightforward to find.
We begin by inverting equation (\ref{real2red}), relating the real 
and redshift displacement fields,
\be
	\xi_i = {\cal P}^{-1}_{ij} \xi^s_j
\label{invdisp}
\ee
where
\be
	{\cal P}^{-1}_{ij} \equiv \delta^K_{ij} - 
	\frac{\beta}{1+\beta} \sh_i \sh_j
\ee
is the inverse redshift projection tensor. Here we have used the 
identity $\sbh=\rbh$.  Taking the divergence of both sides
of equation (\ref{invdisp}) will give us the inverse relationship between 
redshift and real space density perturbations,
\be		
	\delta = - \nabla. \xib = 
	[1- \frac{\beta}{1+\beta}(\de^2_s+
	2s^{-1} \de_s)\nabla^{-2}] \delta^s.
\ee
The inverse spherical redshift operator is given by
\be
	\Sb^{-1} =1-\frac{\beta}{1+\beta}(\de^2_s+2s^{-1}\de_s)\nabla^{-2}
\label{invS}
\ee
To invert the redshift distortion we have used only the continuity equation
in real and redshift space, and assumed potential flow in both the
real and the redshift displacement fields.
 In Section \ref{curl} we shall show that if the linear real space displacement 
field is curl--free then so is the linear redshift displacement field.

Having found the inverse operator we now verify that it
satisfies the relation
\be
	\Sb \Sb^{-1} = \Sb^{-1}\Sb = 1.
\ee
In order to show that equation (\ref{invS}) does satisfy this it is
useful to first find the 
relation between $\Sb$ and ${\cal P}_{ij}$, and $\Sb^{-1}$ and 
${\cal P}^{-1}_{ij}$. From the redshift continuity equation we find
\be
	\delta^s = - \nabla . \xib^s = - \nabla_i {\cal P}_{ij} \xi_j =
 \nabla_i {\cal P}_{ij} \nabla_j \nabla^{-2} \delta.
\ee
Expanding $\delta$ in the same way we find that $\Sb$ and ${\cal P}$
and $\Sb^{-1}$ and ${\cal P}^{-1}$ are related by 
\be
	\Sb \equiv (\nabla_i {\cal P}_{ij} \nabla_j) \nabla^{-2},
	\hspace{1.cm}
	\Sb^{-1} \equiv (\nabla_i {\cal P}_{ij}^{-1} \nabla_j) \nabla^{-2},
\label{defS}
\ee

Multiplying these operators together yields
\ba
	\Sb \Sb^{-1} &=& (\nabla_i {\cal P}_{ij} \nabla_j) \nabla^{-2}
	(\nabla_k {\cal P}_{kl}^{-1} \nabla_l) \nabla^{-2} ,\nn
		&=& \nabla_i {\cal P}_{ij} (\nabla_j \nabla^{-2}
	\nabla_k) {\cal P}_{kl}^{-1} \nabla_l \nabla^{-2}
\label{deriv}
\ea
where we have highlighted the operator $\nabla_j \nabla^{-2}\nabla_k$
in the second line which is applied to the real space displacement 
field, $\xib$.
If we apply this operator to a general potential field $A_i=\nabla_i \phi$
we find
\ba
	\nabla_i \nabla^{-2}\nabla_j A_j &=&
	\nabla_i \nabla^{-2}\nabla_j \nabla_j \phi, \nn
		&=& A_i.
\ea
Hence for a potential vector field this operator is the identity matrix 
\be
	\nabla_i \nabla^{-2}\nabla_j = \delta^K_{ij}. 
\ee

Substituting this into equation ({\ref{deriv}}) completes the proof
\ba
	\Sb \Sb^{-1} &=& \nabla_i {\cal P}_{ij} (\nabla_j \nabla^{-2}
	\nabla_k) {\cal P}_{kl}^{-1} \nabla_l \nabla^{-2} \nn
		&=& \nabla_i {\cal P}_{ij}  
		{\cal P}_{jl}^{-1} \nabla_l \nabla^{-2} \nn
		&=& \nabla_i \delta^K_{il} \nabla_l \nabla^{-2} \nn
		&=& 1.
\ea

\subsection{Is the linear redshift displacement field irrotational ?}
\label{curl}

 To establish the same proof for $\Sb^{-1} \Sb$ we need to show that the operator 
$\nabla_i \nabla^{-2}\nabla_j$, which this time is applied to $\xib^s$,  is acting
on a potential field. Introducing the redshift vorticity vector,
 $\omegab^s$ defined as
\be
	\omegab^s \equiv \beta \nabla \times \xib^s,
\ee
and given that the real space displacement field is curl-free,
\be
	\omegab \equiv \nabla \times \xib =0,
\ee
we find
\ba
	\omegab^s &=& \beta \nabla \times \rbh (\rbh . \xib) \nn
	&=& \beta (\rbh. \xib) (\nabla \times \rbh) + \beta (\rbh \times \nabla)
	(\rbh. \xib) \nn
	&=& 0.
\ea
Hence the linear redshift displacement field is curl--free and our 
 proof holds.

\subsection{Flux limited redshift inversion}
\label{fluxlim}

Since most redshift surveys are flux limited rather than volume limited
it is useful to find the flux limited linear redshift operators. 
The density of galaxies seen in a redshift survey can be represented in 
real space by
\be
	\rho = \phi(1+\delta),
\ee
while in redshift space the galaxy density can be written as
\be
	\rho^s = \phi^s(1+\delta^s).
\ee
The mean observed density of galaxies in the survey is given by
the selection function, $\phi$, in real space and $\phi^s=\phi(\s)$
in redshift space. Conservation of galaxy numbers imply that
\be
	\rho(\r)\, d^3\! r = \rho^s(\s) \,d^3 \!s
\ee
and so  the real and redshift density fields are related by
\be
	\rho = \rho^s \,{\rm det}\, {\cal D}_{ij}
\ee
where 
\be
	{\cal D}_{ij} \equiv \frac{\de s_i}{\de r_j}
\ee
is the distortion tensor mapping real to redshift space.
Expanding the determinant of the distortion tensor to first order we find
\be
	{\rm det} \,{\cal D}_{ij} \approx 1 + 
	\frac{\beta}{1+\beta} \nabla_i \sh_i \sh_j \xi^s_j,
\ee
where we have used the relationship
\be	
	s_i = r_i + \frac{\beta}{1+\beta}\rhat_i \rhat_j \xi^s_j
\ee
to express the expansion in terms of the redshift displacement vector.
We can also expand the real space selection function in terms of redshift space
quantities:
\be
	\frac{\phi^s}{\phi} = 1 - \frac{\beta}{1+\beta} (\sbh .\xib^s)
	\de_s \ln \phi.
\ee
Combining these expressions we find to first order that the 
spherical redshift operator for a flux limited galaxy survey is
\be
	\delta = [1-\frac{\beta}{1+\beta}(\de_s^2 +\alpha'(s) s^{-1} \de_s)
	\nabla^{-2} ] \delta^s.
\ee
where 
\be
	\alpha'(s) \equiv \frac{d\ln s^2 \phi^{-1}(s)}{d \ln s}
\ee
is related to the local slope of the selection function.
Hence the inverse redshift operator for flux limited redshift surveys is
\be
	\Sb^{-1} = 1-\frac{\beta}{1+\beta}(\de_s^2 +\alpha'(s) s^{-1} \de_s)
	\nabla^{-2}.
\ee
Similarly the forward redshift operator is
\be
	\Sb = 1+\beta (\de_r^2 +\alpha(r) r^{-1} \de_r)
	\nabla^{-2}
\ee
where 
\be
	\alpha(r) \equiv \frac{d\ln r^2 \phi(r)}{d \ln r}.
\ee
$\alpha$ and $\alpha'$ are related by 
\be
	\alpha(r) + \alpha'(r) = 4 .
\ee
It is clear from these relations that the effect of changing from
a volume limited redshift survey to a flux-limited redshift survey can
be easily incorporated by the substitution $2 \rightarrow \alpha$
or $2 \rightarrow \alpha'$ in the forward and inverse operators
respectively. This simple transformation will allow us to generalise
results from volume-- to flux--limited.

\subsection{Inverse Redshift Operator in the Local Group Frame}
\label{LGframe}

So far we have only considered the distortion operator and its inverse
from the point of view of an observer at rest with respect to the CMB.
However the true rest frame of an observer is influenced by local,
nonlinear gravitational interactions.  Since nonlinear motions 
induced by local galaxies cannot be treated in the linear 
framework it is usual to transform to the linear part of our
motion, that of the Local Group of galaxies (whether our Local Group
also has a substantial nonlinear component to its motion is an 
issue we shall side-step here). One can then either
work directly in the Local Group frame, or use the known dipole,
measured by the CMB dipole anisotropy and transform to the CMB 
frame. However, there are a number of reasons for working in the
Local Group rest frame. Firstly the linear theory formalism is only valid so 
long as the perturbative variable 
 $\rbh.(\vb(\r)-\vb(\0))/r \rightarrow 0$ when $r \rightarrow 0$. In the CMB
frame the velocity field does not necessarily vanish at the origin,
and an extra, un-necessary assumption is needed to satisfy this.
In addition, if our relative motion does not vanish at the origin,
divergences appear in the analysis of our dipole, calculated from
redshift surveys. We discuss this effect in detail below. Finally,
as we shall show in Section 3 reconstruction of the velocity field 
in the Local Group frame reduces the uncertainties due to finite
survey volumes and shot--noise.

The transform from real space to redshift space in the
Local Group rest frame can be found by subtracting the 
Local Group displacement from the redshift displacement vector:
\be
	\xi^{s,LG}_i={\cal P}_{ij}\xi_j-\beta \rh_i \rh_j 
	\xi_j(\0)
	= {\cal P}_{ij}\xi_j- ({\cal P}_{ij}-\delta^K_{ij})
	\xi_j(\0).
\label{lgdist}
\ee
Super- or sub-script LG denotes a variable measured in the 
Local Group frame, and $\xib(\0)$ is the Local Group displacement, or 
dipole.
This is equivalent to using $\rbh.(\vb-\vb(\0))$ instead of 
$\rbh.\vb$ as the redshift coordinate. From equation (\ref{lgdist}) 
the transformation to redshift space is straightforward. Taking 
the divergence of the redshift
displacement field in the Local Group frame we find
\be
	\delta^s_{\small LG} = 
	[1+\beta(\de_r^2+2r^{-1}(\de_r-\de_{r=0}))\nabla^{-2}]
	\delta
\ee
where $\de_{r=0}$ is the radial differential operator at the origin
and $\de_{s=0}\nabla^{-2}=-\int \! d^3r \, 
(\sbh.\rbh)/4 \pi r^2$ is the radial velocity operator at the origin. 
It is useful to note that $\nabla_i \sh_i \sh_j \xi_j(\0)=
\xi_j(\0)(\nabla_i\sh_j \sh_i)$ since $\xib(\0)$ is a fixed vector at 
the origin. 

In the Local Group frame the redshift operator is then
\be
	\Sb_{\rm LG} = 
	1 + \beta(\de_r^2 + 2 r^{-1} (\de_r -\de_{r=0}))\nabla^{-2}.
\ee
Finding the inverse is slightly trickier. A straightforward inversion 
of equation (\ref{lgdist}) yields
\be
	\xi_i = {\cal P}^{-1}_{ij} \xi^{s,LG}_j + 
	(\delta^K_{ij} - {\cal P}^{-1}_{ij})\xi_j(\0)
\label{invlgdisp}
\ee
Therefore to find the real space displacement field we need to know 
what the Local Group displacement is. One could take the known dipole
from the CMB here, but to calculate the effect self-consistently
the Local Group motion should be estimated from the redshift 
catalogue itself.

One might suppose that equation (\ref{lgdist}) would help us here,
since it transforms real displacements to redshift displacements.
But if we let $r \rightarrow 0$ we find 
$\xib^{s,LG}(\0) \rightarrow \xib(\0)$. This is just a 
statement that the only displacement not affected by the redshift
transformation is the observer's, since this is by definition the 
origin of coordinates in both systems.
But while we have demonstrated that the redshift and real space Local Group 
displacements are equal, this does not imply that the Local Group motion 
estimated from redshift catalogs are the same in real and redshift space,
and it is the latter quantity we need. 
To illustrate this, consider the dipole
generated at the origin. This can be calculated from the density
field by
\be
	\xib(\0) =  \int \! \frac{d^3 r}{4 \pi} \, \frac{\rbh}{r^2} \delta(\r),
\label{realdipole}
\ee
Similarly, one can calculate the redshift dipole (Kaiser \& Lahav 1988)
\ba
	\xib^s(\0) &=&  \int \! \frac{d^3 r}{4 \pi} \, 
	\frac{\rbh}{r^2} \delta^s(\r) \nn
	&=& - \int \! \frac{d^3 r}{4 \pi}\, 
	\frac{\rbh}{r^2} \nabla. \xib^s(\r) \nn
	&=& \xib(\0)\left(\beta \frac{2}{3} \int \! \frac{dr}{r}+1 \right)\nn
	 & & +\beta \int \! \frac{d^3r}{4 \pi} \frac{\rbh}{r^2}
	(\de_r^2 +2 r^{-1}\de_r)\nabla^{-2} \delta,
\ea
which is clearly different from the true dipole. If we only consider
the first term, which arises from the intrinsic dipole motion of
the observer,
\be
	\xib^s(\0) = \beta \xib(\0) \frac{2}{3} \int \! \frac{dr}{r},
\ee
we find a logarithmic contribution to the dipole from the
survey geometry. This is the well known ``Rocket Effect'' 
(Kaiser \& Lahav 1988). This contribution is generated by the reflex 
action of the distorted survey on the measured dipole. 
The divergence of the estimated dipole arises when points in real
space are mapped to the origin in redshift space. In practice
we would not expect this divergence to appear if the velocity 
field is coherent over some scale. In this case the local matter 
field will be moving at the same velocity as the observer and will
not be mapped to the origin. Hence the way to avoid such divergences
is to smooth the velocity field on some scale. In practice we
will have to smooth on large scales to allow the use of linear 
theory (see Section \ref{filter} below). 

A second divergence appears at large radii. As $r\rightarrow \infty$
the Rocket term diverges logarithmically, so the redshift dipole 
in the Local Group frame does not converge (Kaiser \& Lahav 1988). This
consideration shows us that the true dipole can only be estimated over
a finite range of radii. In practice a reconstruction should be done 
for different survey depths to test if the true dipole has converged.

The true dipole can be estimated from redshift space via equation 
(\ref{realdipole}) and the divergence of equation (\ref{invlgdisp})
\ba
	\xib(\0) &=& - \int\! \frac{d^3 r}{4 \pi} \,
	\frac{\rbh}{r^2} \nabla.\xib(\r) \nn
	&=&  \int\! \frac{d^3s}{4 \pi} \frac{\sbh}{s^2} \left(1-
	\frac{\beta}{1+\beta} (\de_s^2 +2 s^{-1}\de_s)\nabla^{-2}
	\right)\delta^{s}_{LG} \nn
	&  & - \frac{2 \beta}{3(1+\beta)}
	\xib(\0)\int\! \frac{dr}{r}.
\ea
Solving this we arrive at the dipole solution
\be
	\xib(\0) =
	 \frac{1}{\cal A} 
	\int\! \frac{d^3s}{4 \pi} \frac{\sbh}{s^2}[1-\frac{\beta}{1+\beta}
	(\de_s^2+2s^{-1}\de_s)\nabla^{-2}]\delta^{s}_{LG},
\label{invdipole}
\ee
where
\be
	{\cal A}= 1+\frac{2\beta}{3(1+\beta)} \int\! \frac{dr}{r}
\ee
The presence of a logarithmic term, $\int\! dr/r$, may seem worrying
but it is there to cancel the divergent term that arises from estimating 
the dipole in redshift space from the Local Group frame. 
Again the divergence at small scale can be removed by smoothing the 
velocity field, or estimating the dipole for a ball of matter at 
the origin. The redshift density field in equation (\ref{invdipole}) 
is acted on by the inverse operator, but no correction is made for the
dipole, so this field is not a mapping to the real space density. We
present the correct expression for this below. In addition, in Section
\ref{secdipole} we show that the dipole is coupled only to the 
redshift space mass dipole. Hence we can estimate the Local Group 
dipole directly from redshift surveys, without
reconstructing the whole survey.

	The inverse redshift distortion operator in the 
local group frame is somewhat more complicated than in the
CMB frame, and can be written as
\be
	\delta = [1-\frac{\beta}{1+\beta}
	(\de_s^2+2s^{-1}\de_s)\nabla^{-2}]
	\delta^s_{LG} - \frac{\beta}{1+\beta} \frac{2}{s} \sbh.\xib(\0),
\ee
where $\xib(\0)$ is estimated from equation 
(\ref{invdipole}), or taken from the CMB dipole.

Writing this in the form of an operator we find that  
the inverse redshift operator in the Local Group frame is 
\ba
	\Sb_{\small LG}^{-1} &=& \left(1+\frac{\beta}{1+\beta}
	\frac{2}{{\cal A}s} \de_{s=0}\nabla^{-2} \right) \nn
	& & \times
	\left(1-\frac{\beta}{1+\beta}(\de_s^2+2s^{-1}\de_s)\nabla^{-2}
	\right).
\ea

These operators are valid for volume--limited redshift surveys.
For flux limited surveys the effects of the selection function are 
easily incorporated by making the substitution $2 \rightarrow \alpha(r)$
in the operator $\Sb$ and substituting $2 \rightarrow \alpha'(s)$ in 
the operator $\Sb^{-1}$, and noting that these are now functions of 
radius and so should appear to the right of the integral.
This completes our task of finding the 
linear inverse redshift distortion operators.

\section{Spherical Harmonic representation of the Redshift Operators}
\label{sphharm}

\subsection{The redshift space operators}

 In the case of
spherical survey geometry and radial distortions the most 
convenient basis is that of spherical harmonics (Heavens \& Taylor 1995,
Ballinger at al 1995, Fisher at al 1995, Tadros et al 1999).
In this basis we can use the properties of the eigenvalues of the
Laplacian to transform from an integro-differential operator to 
an algebraic operator. We shall assume for convenience that the survey
is all sky. For a spherical harmonic function, 
$\psi_{\ell m}(k)$, defined by 
\be
	\psi_{\ell m}(k) = \sqrt{\frac{2}{\pi}}
	\int\! d^3 r \,\psi(\r)j_\ell(kr)Y^*_{\ell m}(\rbh)
\ee
the Laplacian becomes 
\be
	\nabla^2 \psi = 
	\left(\de_r^2 + \frac{2}{r} \de_r - \frac{L^2}{r^2}\right) \psi
	= -k^2 \psi
\ee
where $L^2\equiv\ell(\ell+1)$ is the square of the angular momentum operator. 
In this basis we can rewrite the distortion operator as
\be
	\Sb = 1 + \beta \left(1 - \frac{L^2}{k^2 r^2} \right)
\ee
whose inverse is
\be
	\Sb^{-1} = 1 - \frac{\beta}{1+\beta} 
	\left(1 - \frac{L^2}{k^2 s^2} \right).
\ee
These operators contain both wavemodes and radial coordinates and
should be used in expressions such as
\be
	\delta(\r) = \sqrt{\frac{2}{\pi}}
	\int\! dk k^2 \sum_{\ell m}\delta^s_{\ell m}(k) \Sb^{-1}(k,r) 
	j_\ell(kr)Y^*_{\ell m}(\rbh),
\label{invkr}
\ee
transforming real space harmonic modes into the redshift space 
density field.
The physical interpretation of these operators is straightforward.
For $\Sb$ an isotropic term proportional to $\beta$ is
added to the true field and the angular part proportional to $L^2$
is subtracted, leaving only a radial distortion. The inverse operator, 
$\Sb^{-1}$, replaces the angular 
term, producing an isotropic distortion, and then reduces the field by 
a factor $1/(1+\beta)$.

To including the effects of a selection function we add an
extra term
\be
	\Sb = 1 + \beta \left(1 - \frac{L^2}{k^2 r^2} \right)
	+ \beta  \frac{\alpha-2}{r} \de_r \nabla^{-2},
\ee
where the integro-differential term can be replaced using the recurrence relation 
for spherical Bessel functions:
\be
	\frac{1}{r}\de_r \nabla^{-2} j_\ell(kr)Y_{\ell m}(\rbh) = 
	- \frac{(\ell j_{\ell-1}(kr)-
	(\ell+1) j_{\ell+1}(kr))}{(2\ell+1)kr} Y_{\ell m}(\rbh),
\ee
causing some mixing between $\ell$-modes.
Moving to the Local Group frame requires another extra term
\be
	\Sb = 1 + \beta \left(1 - \frac{L^2}{k^2 r^2} \right)
	+ \frac{\beta}{\sqrt{3 \pi}kr} \delta^K_{\ell 1}
	\delta^K_{m 0} \int\! d ^3r' \, \delta_D(\r')
\ee
where the last term is the dipole contribution in spherical 
harmonics (see section \ref{dipolesection}), and the integral operator 
acts to shift positions to the origin.

\subsection{The transformation properties of harmonic modes}

Another representation is the transformation of the spherical 
harmonic modes of the redshift survey. In this case the modes
transform according to the relation
\be
	\delta^s_{\ell m}(k) = (1+\beta) \delta_{\ell m}(k)
	 - \frac{\ell(\ell+1)}{2\ell+1} 
	\beta \int^\infty_0 \!\! dk' \,\delta_{\ell m}(k')
	\kappa_\ell(k',k)
\label{CMB_spherical}
\ee
where we have used the integral relation (Watson 1966)
\be
	\int_0^\infty\! dr \,j_\ell(kr) j_\ell(k'r) = \frac{\pi}{2(2\ell+1)}
	\kappa_\ell(k',k)
\ee
and the radial function $\kappa_\ell(k',k)$ is defined as
\ba
\kappa_\ell(z,z') &\equiv&\frac{z'^\ell}{z^{\ell+1}},\hspace{2.cm} z\ge z' \nn
		       & &\frac{z^\ell}{z'^{\ell+1}}, \hspace{2.cm} z'>z.
\label{kappa}
\ea
The inverse transformation, from $\delta^s_{\ell m}(k)$ to $\delta_{\ell m}(k)$,
is done in the same way, with the substitutions 
$(1+\beta)\rightarrow (1+\beta)^{-1}$
in the first term in equation (\ref{CMB_spherical}), and $\beta \rightarrow
-\beta/(1+\beta)$ in the second.

Similarly the redshift transformation in the Local Group frame
is 
\ba
	\lefteqn{\delta^{s,LG}_{\ell m}(k) = (1+\beta) \delta_{\ell m}(k)
	 - \frac{\ell(1+\ell)}{2\ell+1} 
	\beta \int^\infty_0\! dk' \,\delta_{\ell m}(k')
	\kappa_\ell(k',k)}\nn 
	& & + \frac{2}{3} \beta \delta^K_{\ell 1}
	\int^\infty_0\! dk'\, \delta_{\ell m}(k') (k'/k^2)
\label{LG_spherical}
\ea
where the radial transform of the dipole term is completed by use 
of the integral (Watson 1966)
\be
	\int_0^\infty\! r dr \, j_\ell(kr) = \frac{\sqrt{\pi}}{k^2} 
	\frac{\Gamma[(\ell+2)/2]}{\Gamma[(\ell+1)/2]}
\ee

\begin{figure}
\vspace{0.cm}
 \centerline{\epsfig{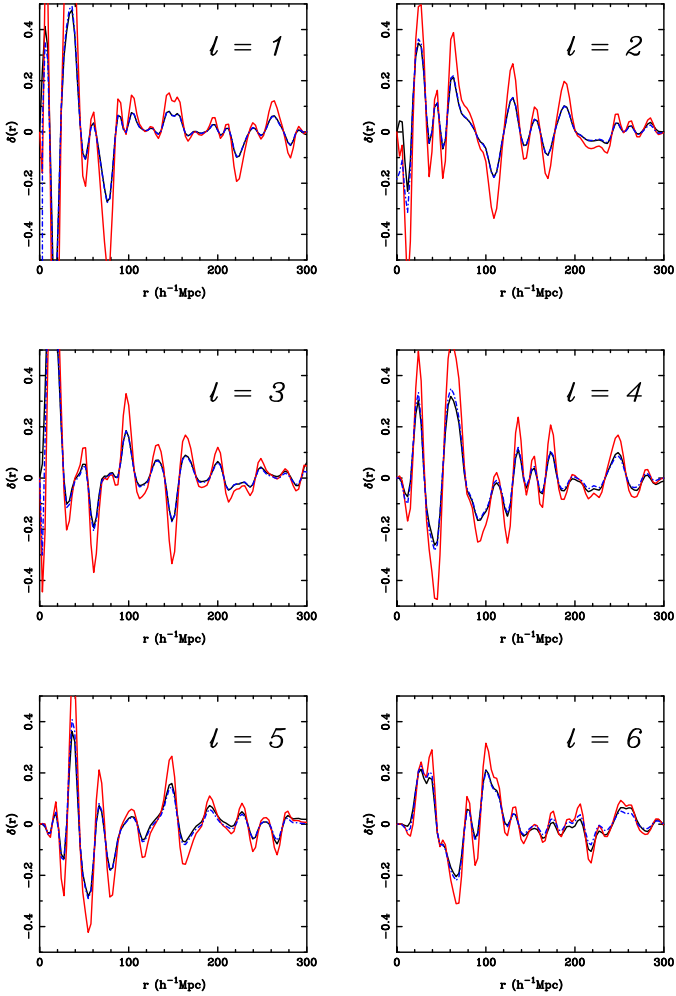}}
\vspace{0.cm}
 \caption{$\ell=1$ to $6$ and $m=0$ density wave in real space (black line),
redshift space (light-grey line) and recovered density field (dark--grey,
dotted line) in the CMB frame with $\beta=1$.
\label{re_density}}
\end{figure}

In Figure \ref{re_density} we show set of Gaussian density fields, 
$\delta_{\ell m}(r)$ in real and redshift 
space, using the transformation equation (\ref{CMB_spherical}). 
The harmonic modes of a constrained Gaussian field 
can be generated by the relation
\be
	\delta_{\ell m}(k) = \sqrt{P(k)} e^{i \theta_{\ell m}(k)}
\ee
where $P(k)$ is the linear power spectrum derived by Peacock \& Dodds
(1994) and $\theta_{\ell m}(k)$ is randomly distributed for
each mode. The waves
have $\ell=1$ to $6$ and $m=0$. Since the angular
modes are independent and the distortion is $m$-independent this choice
of azimuthal mode does not affect this demonstration. The 
real space density wave was recovered using equation
(\ref{invkr}). It is clear in this test of linear reconstruction
that modes can be recovered with great accuracy.

\subsection{The redshift power spectrum}

Having calculated the harmonic modes of the density field in redshift 
space, we can calculate their correlations. This has been calculated before
by Heavens \& Taylor (1995) for a discrete set of spherical harmonics 
and by Zaroubi \& Hoffmann (1996) for a set of Fourier harmonics. 
However the methods presented here provide a simpler expression. 

Taking the two--point expectation value of the harmonic modes we
find
\ba
	\lefteqn{\lgl \delta^s_{\ell m}(k)\delta^{s*}_{\ell m}(k') \rgl = 
	(1+\beta)^2 P(k) k^{-2} \delta_D(k-k')} \nn 
	& & + 
	\beta^2 \frac{\ell^2(\ell+1)^2}{(2\ell+1)^2} \int^\infty_0 \!
	dk_1 P(k_1) k_1^{-2} \kappa_\ell (k_1,k) \kappa_\ell (k_1,k')\nn
	& & - 
	\beta (1+\beta) \frac{\ell(\ell+1)}{2\ell+1}
	(P(k) k^{-2} \kappa_\ell (k',k)+P(k') k'^{-2} \kappa_\ell (k,k')),\nn
\ea
and zero for different $\ell$'s or $m$s.
As well as redshift space distortions creating correlations between
different modes, the second term shows that the shape of the redshift
power spectrum depends on the overall shape of the real power spectrum 
through a convolution. 


\subsection{The cosmological dipole}
\label{secdipole}

In Section 2.4 we derived an expression for the real space
Local Group dipole. Again this takes a simple form in 
spherical harmonics. Expanding $\delta^s$ in harmonics, and 
using the relation
\be
	\rh_i = \sqrt{\frac{4 \pi}{3}} Y_{1 i}(\rbh)
\ee
and the orthogonality of spherical harmonics, we can reduce 
equation (\ref{invdipole}) to
\be
	\xi_i(\0) = \frac{1}{\sqrt{6 \pi^2}{\cal A}}
	\int^\infty_0\!\! dk k^2 \int\! dr \, 
	\Sb^{-1}_{\ell=1}(k,r) \delta^s_{1 i}(k) j_1(k r)
\label{dipoleharm}
\ee
The radial integral can be performed to give
\ba
	\xi_i(\0) &=&  \frac{1}{\sqrt{6 \pi^2}{\cal A}(1+\beta)}
	\int^\infty_0 \!\!dk k \, \delta^s_{1 i}(k) \nn
	& & \times [j_0(t)+ 2/3 \beta 
	(j_0(t)+j_1(t)/t-{\rm Ci}(t))]^{kr_0}_{kR}
\label{spheredipole}
\ea
where $R$ and $r_0$ are the upper and lower radial limits of the survey
and ${\rm Ci}(z)=-\int_z dt \cos(t)/t$ is the cosine integral.

Alternatively one can express the dipole in terms of the redshift 
space density field without reference to the harmonic expansion.
Integrating equation (\ref{dipoleharm}) and using the definition 
of the density dipole we find the real space dipole reduces to
\be
	\xib(\0) = \frac{1}{{\cal A}(1+\beta)}
	\int \! \frac{d^3\!s}{4 \pi} \, \delta^s(\s) \sbh 
	  \left( \frac{1}{s^2} +\frac{\beta}{3}\left(
	\frac{1}{s^2} + \frac{1}{R^2} - \frac{2}{r_0 s} \right) \right)
\label{dipolereal}
\ee
 Equation (\ref{spheredipole}) or (\ref{dipolereal}) can also be used to 
estimate the real space dipole contribution in shells. In the limit
of no distortions ($\beta\rightarrow 0$) equation (\ref{dipolereal}) reduces
to the real space dipole, given by equation (\ref{realdipole}), and 
equation (\ref{spheredipole}) reduces to the harmonic equation for the 
dipole (equation (\ref{dipole}), Section \ref{dipolesection}).

Hence we see that the real space velocity dipole can be reconstructed
from the redshift space density field, without a full reconstruction.
Added complications arise if there is incomplete sky coverage, since 
modes will be mixed, and this mixing must be included in a dipole 
estimate. 

Finally, we can add on the effects of a selection function. Making the 
substitutions $2 \rightarrow \alpha'$ in equation 
(\ref{invdipole}) and working through, we find the selection function
correction term to equation (\ref{dipolereal}) is 
\be
	\xi^{\phi}_i(\0) = \frac{1}{12 \pi  {\cal A}(1+\beta)}
	\int \! d^3\!s \, \delta^s(\s) \sbh \int^R_{s} \! \frac{ds'}{s'^2} \, 
	\de_s \ln \phi(s').
\ee
where the factor of $2$ in ${\cal A}$ is also transformed to $\alpha'$.

\subsection{Simulated reconstruction of the dipole motion}

\begin{figure}
\vspace{-3.cm}
 \centerline{\epsfig{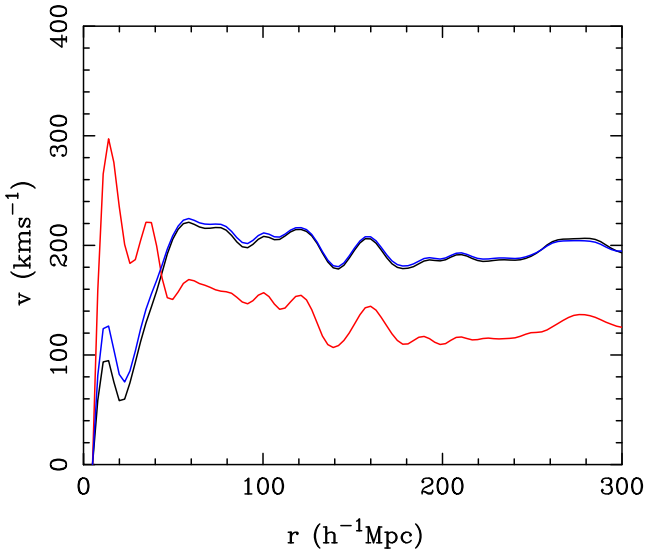}}
\vspace{-3.cm}
 \caption{True CMB redshift and recovered dipole from a random Gaussian
realisation of a density field, calculated  in the CMB rest frame.
Only the amplitude of the dipole is plotted. The light grey line is 
the redshift space dipole. The darker lines are the true and reconstructed
dipoles. The fields  have been
Gaussian smoothed on a scale of $5h^{-1}{\rm Mpc}$.
\label{re_dipole_CMB}}
\end{figure}


Figure \ref{re_dipole_CMB} shows the amplitude of the 
dipole reconstructed from a
simple random Gaussian model of the density field.
 The true dipole was calculated from equation 
(\ref{spheredipole}) with $\beta=0$ (black line). Transforming the spherical 
harmonic modes to redshift space in the CMB frame was then done via equation 
(\ref{CMB_spherical}), with $\beta=1$, and the dipole re-calculated in 
redshift space (grey line). Finally the true dipole (dark grey line) was 
recovered via equation (\ref{spheredipole}). 
It is clear from figure \ref{re_dipole_CMB} that 
the true dipole can also be recovered to good accuracy from equation 
(\ref{spheredipole}).

\subsection{Filtering redshift surveys}
\label{filter}

A number of methods for reconstructing the density and 
velocity fields do not use a smoothed galaxy distribution, but rather
treat the galaxies in the catalog as test particles, inverse
weighted by the selection function. The question then arises,
should one smooth or not? Clearly information is lost whenever
one smoothes. However, there are a number of reasons why one 
can smooth in reconstruction. Firstly we are applying linear
theory to the density field, so the scales we are considering must be 
large. Secondly, as well as nonlinear features in real space,
there are nonlinear caustics and ``fingers of god'' in redshift space
that we must remove before applying the present formalism. An alternative
to smoothing ``fingers of god'' is to collapse them back onto the 
cluster they originate from. The small distances between galaxies in
clusters can also lead to divergences in the calculated velocity field
when the distance between galaxies becomes small. Hence smoothing
would seem to be a good thing.

However smoothing in redshift space is not the same as smoothing in
real space, since the coordinate system changes between the two. 
An isotropic smoothing kernel in one space will not be isotropic in the other.
Anisotropic smoothing kernels, such as in adaptive, or 
Lagrangian smoothing which uses the local moment of interia as
a kernel, will preserve the shape of structure and can
be mapped back to real space smoothed with the correct kernel.
  
The spherical harmonic decomposition uses sharp $k$- and $(\ell,m)$-space 
filtering. This has the advantage that both redshift and real space will have the 
same smoothing, although since the harmonics are related by a 
convolution over $k$-space we need to sample higher $k$-modes in the 
inverted space. This will be limited by nonlinearities in the higher 
$k$ range. The sharpness of the $k$-space filter can be removed by introducing 
extra filters.

\section{Properties of reconstructed redshift surveys}
\label{prop}

Having found the inverse redshift space operator, we can now 
transform redshift surveys from redshift space to real space,
and in the process predict the linear cosmological velocity field
and the Local
Group dipole. In doing so it is useful to have some idea of what
uncertainties arise in the process. In this Section we calculate 
some of the properties of the reconstructed
density and velocity fields. Throughout we shall assume that the
survey is spherically symmetric, which is nearly the case for the IRAS
Point Source Redshift Survey (Saunders et al 1996).

It is useful to disentangle the various uncertainties in 
analysing redshift surveys. In the following sections we treat
independently the effects of:  a biased and uncertain input 
distortion parameter, $\beta$ (Section \ref{wrongbeta}); a finite survey 
volume on the reconstructed density field (Section \ref{cosvarden});
the properties of the velocity field in the CMB frame for finite survey volumes 
(Section \ref{cosvarvel}); a finite survey on the dipole (Section \ref{cosvardi});
the properties of the velocity field in the Local Group frame 
(Section \ref{dipolesection}); shot-noise (Section \ref{sncont}).
Apart from Sections \ref{wrongbeta} and \ref{cosvarden} we shall ignore 
the effects of the redshift distortion. 
While this over-simplifies the case, it is useful 
to understand each of the effects in isolation.

\subsection{The wrong $\beta$.}
\label{wrongbeta}

Using the incorrect distortion parameter, $\beta$, will lead 
to a systematic offset in the recovered fields. If the true
distortion is $\beta$, and we use a different value $\beta'$ in the
reconstruction, a residual term is left in the reconstructed density
field. This can be characterised by finding the product $S'^{-1}S$,
where $S'^{-1}$ is the inverse operator with the incorrect distortion parameter.
This residual is
\be
	S'^{-1}S = 1 + \frac{\beta-\beta'}{1+\beta'} 
	(\de_r^2 + 2r^{-1}\de_r) \nabla^{-2},
\ee
for a constant offset distortion. Another possibility is that there
is a large uncertainty on the distortion parameter (as is the case for
the present generation of redshift surveys) and a stochastic 
term is added to the reconstructed fields. If we now take $\beta'$ to
be an unbiased but uncertain estimate of the true distortion parameter,
the induced uncertainty on the density field is
\be
	\sigma_{\delta'}  = 
	\frac{\sigma_{\beta}}{(1+\beta)^2} (\de_r^2 + 2r^{-1}
	\de_r) \nabla^{-2} \delta .
\ee
Hence the uncertainty due to errors in the distortion parameter 
lead to a density--dependent uncertainty in the recovered fields.

\subsection{Incompleteness in reconstructed density fields}
\label{cosvarden}

The reconstruction of the cosmological density field is always
incomplete, since the distortion is nonlocal and we are trying
to reconstruct with only a finite volume. Hence uncertainties 
arise in our estimates of the velocity and shear fields which 
cause the distortion. We
can quantify this uncertainty in a model-dependent fashion by
calculating the effects of a finite survey and finding the 
rms contribution from external structure not included in the
survey volume.

The effect of a finite survey volume is to alter the true inverse
redshift operator, $S^{-1}$, to an effective one, $S_{\rm eff}^{-1}$,
using only the information within the surveyed region. Using the 
definition of the inverse redshift operator (equation \ref{defS})
we see that this is equivalent to the operator
\be
	S_{\rm eff}^{-1} 
	\equiv (\nabla_i {\cal P}_{ij}^{-1} \nabla_j) \nabla^{-2}_{\rm eff}
\ee
where $\nabla^{-2}_{\rm eff}$ is the inverse Laplacian defined 
over a finite volume. It is useful to split the inverse Laplacian defined
over all space, $\nabla^{-2}$, into a term defined within the survey volume,
$\nabla^{-2}_{<R}$, and a term defined outside the survey volume, 
$\nabla^{-2}_{>R}$,
\be
	\nabla^{-2}=\nabla^{-2}_{<R}+\nabla^{-2}_{>R}.
\ee
By doing so we can find the difference between the reconstructed 
density field and the true real space density field in terms of 
$\nabla^{-2}_{>R}$. Since our effective inverse Laplacian is equivalent
to $\nabla^{-2}_{<R}$, we can write 
\be
	\nabla^{-2}_{\rm eff} = \nabla^{-2} - \nabla^{-2}_{<R}.
\ee

Let us define
\be
	\Delta \equiv [S_{\rm eff}^{-1} S] \delta - \delta
\ee
as the residual field after reconstruction. We find that this field 
can be expressed as
\be
	\Delta = - [\Sb^{-1} \nabla^2 \nabla^{-2}_{>R} \Sb] \, \delta.
\ee
Expanding the density field in spherical harmonics and applying these 
operators we find
\be
	\Delta(\r)=\sqrt{\frac{2}{\pi}}
	\frac{\beta}{1+\beta}\int \!dk k^2  \sum_{\ell m} 
	\delta_{\ell m}(k) 
	\frac{\ell(\ell+1)}{2\ell+1} 
			 \frac{F_\ell(k,R)}{(kr)^{2-\ell}}
			Y_{\ell m}(\rbh) 
\ee
where 
\be
	F_\ell(k,R) = \int_R^\infty \! \! \frac{dr'}{r'} (kr')^2
	\left( 1+\beta - \beta \frac{\ell(\ell+1)}{k^2 r'^2}\right) 
	\frac{j_\ell(k r')}{(kr')^\ell}. 
\ee
For $\beta=0$ we have the trivial solution that the residual field
is zero, since no reconstruction is necessary. We also note that the
monopole mode is zero from 
Newtons' First Theorem. The $\ell=1$ dipole term increases with decreasing 
radius and diverges at the origin,
suggesting that this should be removed in reconstructions by working in the
Local Group frame (we shall return to this point when reconstruction
velocity fields), while the $\ell=2$ quadrupole residual is independent of radius.

The variance of these terms can be found by squaring and ensemble
averaging using the result that the ensemble average of the 
$\delta_{\ell m}(k)$'s
is (Heavens \& Taylor 1997)
\be
	\lgl \delta_{\ell m}(k)\delta^*_{\ell' m'}(k') \rgl =
	P(k) k^{-2} \delta_D(k-k') \delta^K_{\ell \ell'} \delta^K_{m m'}
\label{power}
\ee
where $\delta_D(x)$ is the Dirac delta function and $\delta^K_{i j}$ is the
Kronecker delta.
The cosmological variance introduced by incompleteness is given by
\ba
	\lgl \Delta^2(r) \rgl = 
	\left[\frac{\beta}{1+\beta}\right]^2
	\! \int_0^\infty \! \! \frac{k^2 dk}{2\pi^2}  
	P(k)  \sum_\ell \frac{\ell^2(\ell+1)^2}{2\ell+1} 
	\left[\frac{F_\ell(k,R)}{(kr)^{2-\ell}}\right]^2 . \nn
\ea

\begin{figure}
\vspace{-3.cm}
 \centerline{\epsfig{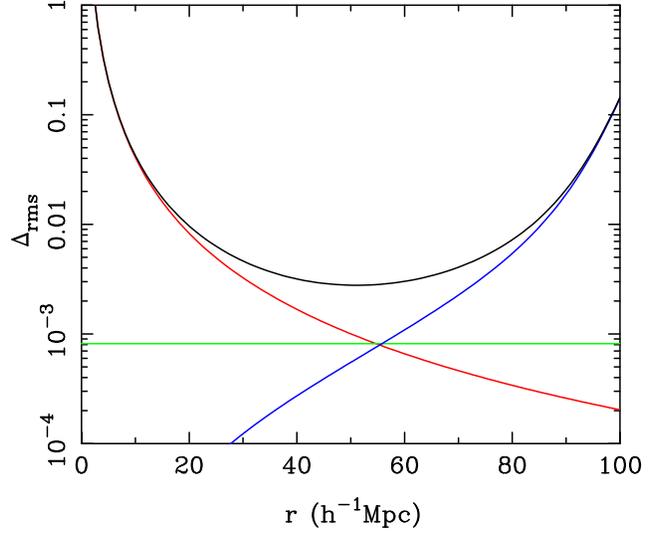}}
\vspace{-3.cm}
 \caption{Cosmic variance in reconstructed density field due to 
 external structure for a volume--limited redshift survey with 
 radius $R=100\hMpc$.
 The upper line is the total uncertainty from all modes. The 
 line decreasing with radius is the $\ell=1$ dipole contribution. This
diverges at small radius. The flat light grey line is the $\ell=2$
 quadrupole contribution. The increasing line is the total contribution
 from all modes $\ell \ge 3$.
\label{density_residual}}
\end{figure}

Figure \ref{density_residual} shows the radial dependence of the rms of 
this residual field for $\beta=1$ and a maximum radius of
 $100 \hMpc$. We again use the Peacock--Dodds form for the power 
spectrum. The upper dark line is the total rms residual
field. The overall impression is that for reconstructing density 
fields the uncertainty is low, except at the center and edges 
of the survey. Linear theory then suggests that density fields can
be recovered to high precision, ignoring the effects of a selection
function and  shot noise.
To understand where this uncertainty comes from we have also plotted
the rms residuals for the dipole and quadrupole. As expected the dipole 
increases at the origin, while the quadrupole is uniform across the 
reconstructed survey. The higher modes, $\ell \ge 3$ contribute to 
a term increasing with radius, but which does not diverge at the survey
boundary. Interestingly the dipole, quadrupole and the contribution from
the rest of the multipoles are all equal at about half the survey radius,
independent of the survey radius. Thus a nice way to characterise the
uncertainty is in terms of the quadrupole contribution, which is independent
of radii, and a third of minimum uncertainty.

\subsection{Incompleteness in the velocity field}
\label{cosvarvel}

The incompleteness that arises when reconstructing the density 
field is not the only worry when calculating cosmic fields from 
a finite region. In the rest of this  Section we consider the effect of a
finite volume on estimates of the velocity field 
in the CMB and Local Group frames. To simplify the analysis we shall
 assume no distortion, and only consider
the uncertainty due to a finite survey volume. We shall also consider 
the effects of shot noise on the reconstructed velocity field. We
begin by expanding the velocity field in spherical harmonics.

The Newtonian Green's function in equation (\ref{eqvel}) can be expanded 
in spherical harmonics
\be
	\frac{1}{|\r'-\r|} = 
	4 \pi \sum_{\ell m} \frac{1}{2\ell+1} \kappa_\ell(r,r') 
	Y_{\ell m}(\rbh)
	Y_{\ell m}^*(\rbh')
\label{newtgreen}
\ee
and is related to the velocity field by $\nabla (1/r)=-\rbh/r^2$. The 
radial function $\kappa_\ell(r,r')$ is given by equation (\ref{kappa})

The gradient operator can be conveniently decomposed into radial and 
transverse terms;
\be
	\nabla [\kappa_\ell  Y_{\ell m}(\rbh)] =
	[\de_r \kappa_\ell] \Y^{L}_{\ell m}(\rbh) -
	i \sqrt{\ell(\ell+1))} \kappa_\ell r^{-1} \Y^{M}_{\ell m}(\rbh)
\ee
where we have defined the vector spherical harmonics 
\be
	\Y^L_{\ell m} = \rbh Y_{\ell m}, \hspace{1.cm}
	\Y^M_{\ell m} = \frac{1}{\sqrt{\ell(\ell+1)}}(\rbh \times \Lb) 
	Y_{\ell m}.
\ee
and $\Lb\equiv-i \r \times \nabla$ is the classical angular momentum operator.
With these definitions we find that 
the radial  velocity component is given by 
\be
	v_r(\r) = \lmdk \delta_{\ell m}(k) U_{\ell m}(k,\r)
\ee
where  
\be
	U_{\ell m}(k,\r) = \frac{1}{2\ell+1} \int_R dr' r'^2 
	[\de_r \kappa_\ell(r,r')] j_\ell(kr') \Ylm(\rbh).
\ee
The  transverse velocity components are given by 
\be
		\vb_t(\r) = \lmdk \delta_{\ell m}(k)\Wb_{\ell m}(k,\r) 
\ee
where the kernel is a transverse vector field; 
\be
	\Wb_{\ell m}(k,\r)= - \frac{i \sqrt{\ell(\ell+1)}}{(2\ell+1) r} 	
	\int_R \! dr' r'^2 \, \kappa_\ell(r,r') j_\ell(kr') 
	\Y^M_{\ell m}(\rbh).
\ee

Using equation (\ref{power}) for the expectation value of two 
harmonic modes we find the variance of the velocity field in terms
of the power spectrum is
\be
	\lgl v^2_r(r) \rgl = \int_0^\infty \!\!\frac{k^2 dk}{2 \pi^2} P(k)
	|U_{\ell m}(k,r)|^2,
\ee
where the effective window function is 
\be
	|U_{\ell m}(k,r)|^2 = 
	\sum_\ell \frac{\ell^2}{(2 \ell+1)} \left[\int_R dr' r'^2 
		[\de_r \kappa_\ell(r,r')] j_\ell(kr')\right]^2
\label{Uwindow}
\ee
and 
\be
	\lgl \vb^2_t(r) \rgl = \frac{1}{r^2} 
	\int_0^\infty \!\!\frac{k^2 dk}{2 \pi^2} P(k)
	|W_{\ell m}(k,r)|^2,
\ee
with the window function
\be
	|W_{\ell m}(k,r)|^2 =
	\sum_\ell \frac{\ell(\ell+1)}{(2 \ell+1)}  \left[\int_R dr' r'^2 
		\kappa_\ell(r,r') j_\ell(kr')\right]^{2}.
\label{Wwindow}
\ee

The range of these radial integrals in the window functions determines the 
source of the 
uncertainties. One of the main uncertainties in reconstruction is
the effect of the density field external to the survey volume.
 In the next Section we study the effects of structure
beyond the survey.  These results can also be
used to calculate the expected variance in the velocity field 
from shells or bulk motions.

\subsubsection{Incompleteness due to external structures}

The range of the radial integral in these equations depends on the 
region of space 
under consideration. If we are determining the sampling variance due to 
structures external to the redshift survey, then the range is $r>R$, where 
$R$ is the radius of the survey (here we shall only consider
sharp edges to the survey. A more complete analysis will include weighting 
and the effects of an angular mask). In this case $r\leq r'$ and so 
$\kappa_\ell=r^\ell/r'^{\ell+1}$. Hence the radial integral in 
equations (\ref{Uwindow})  and (\ref{Wwindow}) is
\be
	\int_R^\infty dr' r'^2 \kappa_\ell(r,r') j_\ell(kr')
 	= k^{-2} (kr)^\ell \left( \frac{j_{\ell-1}(kR)}{(kR)^{\ell-1}}\right)
\ee
The radial derivative with respect to $r$, in equation (\ref{Uwindow}) is trivial.

The total sampling variance due to external structure is then
\be
	\lgl v_{\rm cv}^2(r) \rgl = \int_0^\infty\!\! \frac{dk}{2 \pi^2} P(k) 	
	\sum_{\ell=1}^\infty \ell (r/R)^{2(\ell-1)} j^2_{\ell-1}(kR)
\ee
The summation is only non-zero from $\ell=1$, since external 
structure cannot affect the monopole mode.

\begin{figure}
\vspace{-3.cm}
 \centerline{\epsfig{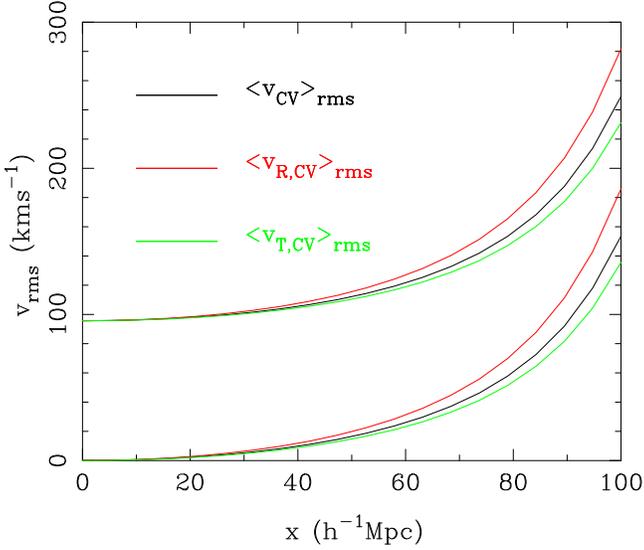}}
\vspace{-3.cm}
 \caption{Cosmic variance in reconstructed velocity field due to 
structure external to a redshift survey with radius $R=100h^{-1}$Mpc.
The upper set of three lines are the radial, total and transverse rms 
uncertainty in the velocity field and its components in the CMB rest 
frame. The lower set of lines are for the same survey, but calculated in the
Local Group rest frame. 
\label{velext}}
\end{figure}

Figure \ref{velext} shows the radial, transverse and total rms velocities, 
$v_{\rm rms}=\sqrt{\lgl v^2 \rgl}$, for a survey with 
$R=100\hMpc$ in the CMB and Local Group frames (see Section 
\ref{dipolesection}) frame. The upper set of three lines correspond
to the radial, total and transverse rms uncertainties in the velocity
field and its components in the CMB rest frame. Interestingly, for a 
spherical survey the uncertainties obey the relation 
$\lgl \vb^2_t(r) \rgl<\lgl v_{\rm cv}^2(r) \rgl<\lgl v^2_r(r) \rgl$. At the
origin the uncertainty is about $100 {\rm kms}^{-1}$, which is the dipole
uncertainty for a survey of radius $R=100\hMpc$ (see Section \ref{cosvardi}).
The main effect is a rise in uncertainty from the center of the survey to
the edges. However, even on the survey boundary the error is finite.

The radial and transverse velocity cross--correlation functions are 
\ba
	\lefteqn{\lgl v_r(\r) v_r(\r') \rgl =}\nn
	& &  \int_0^\infty \!\! \frac{dk}{2 \pi^2} P(k) 
	\sum_\ell \frac{\ell^2}{(2 \ell+1)} 
	\left(\frac{rr'}{R^2}\right)^{\ell-1} \!\! j^2_{\ell-1}(kR)
	{\cal P}_\ell(\mu)
\ea 
and 
\ba
	\lefteqn{\lgl v_t(\r) v_t(\r') \rgl =}\nn
	& &  \frac{1}{r^2} \int_0^\infty \!\! \frac{dk}{2 \pi^2} P(k) 
	\sum_\ell \frac{\ell(\ell+1)}{(2 \ell+1)} 	
	\left(\frac{rr'}{R^2}\right)^{\ell-1} \!\! j^2_{\ell-1}(kR)
	{\cal P}_\ell(\mu)
\ea
where ${\cal P}_\ell(\mu)$ is the Legendre function and $\mu=\rbh.\rbh'$.
No cross terms arise in the case of an all--sky survey, as we have 
chosen orthogonal projections.

\subsubsection{Sampling variance from interior structure}

Over the range $r<R$ the radial integrals in equations (\ref{Uwindow})
and (\ref{Wwindow}) can be again calculated;
\ba
	\lefteqn{\int_0^R dr' r'^2 \kappa_\ell(r,r') j_\ell(kr')
 	=}\nn
	& & \left(\frac{2 \ell+1}{k^2}\right) j_\ell(kr) - 
	\frac{(kr)^\ell}{k^2}  
	\frac{j_{\ell-1}(kR)}{(kR)^{\ell-1}}
\ea
Differentiating with respect to $r$ yields
\be
	  \left(\frac{2 \ell+1}{k}\right) j'_\ell(kr) - 
	\frac{\ell}{k}  \left(\frac{r}{R}\right)^{\ell-1}j_{\ell-1}(kR)
\ee
In the limit that $R\rightarrow \infty$ we recover the expression 
for the true peculiar velocity field (Reg\"{o}s \& Szalay 1989)
\be
	v_r(\r) = \lmdk \delta_{\ell m}(k) k^{-1} j'_\ell(kr) Y_{\ell m}(\rbh)
\ee
where we a dash on the spherical Bessel function denotes differentiation 
with respect to the argument. The autocorrelation function of the radial
velocities is 
\be
	\lgl v_r(\r) v_r(\r') \rgl =   \int_0^\infty \!\! \frac{dk}{2 \pi^2}
	P(k) \sum_\ell (2\ell+1) j'_\ell(kr) j'_\ell(kr') {\cal P}_\ell(\mu)
\ee
and their variance is
\be
	\lgl v^2_r(r) \rgl =  \int_0^\infty \!\! \frac{dk}{2 \pi^2}
	P(k) \sum_\ell (2\ell+1) j^{'2}_\ell(kr). 
\ee

\subsection{Incompleteness in the cosmological dipole}
\label{cosvardi}

As $r\rightarrow 0$ only the $\ell=1$ (dipole) term survives and we 
recover the cosmic variance on the dipole
\be
		\lgl v_{\rm cv}^2(0) \rgl =
	  \int_0^\infty \frac{dk}{2 \pi^2} P(k) j^2_{0}(kR).
\ee

\begin{figure}
\vspace{-3.cm}
 \centerline{\epsfig{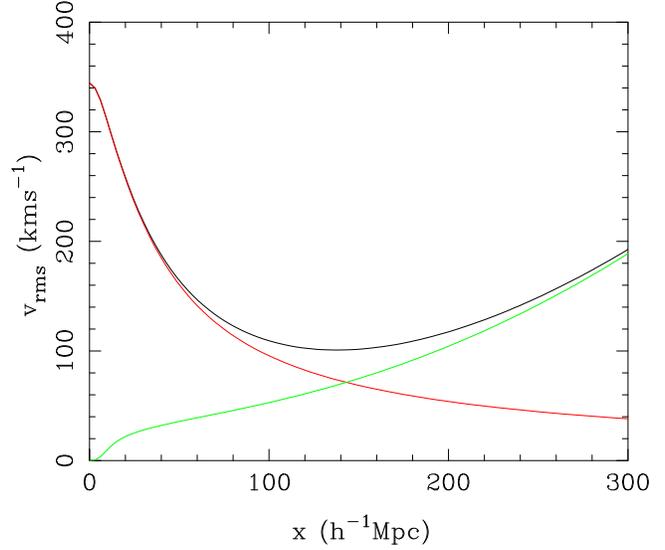}}
\vspace{-3.cm}
 \caption{Uncertainty in the reconstructed dipole due to cosmic variance 
from external structure (dark line) and shot noise due to the discrete
sampling of galaxies in the survey (light line; see Section \ref{sncont}). 
The upper
curve is the total rms uncertainty due to cosmic variance and shot noise 
added in quadrature. The power spectrum is described in the text and 
the redshift survey selection 
function is modelled on the PSCz  (Saunders et al 1996).
\label{dipolefig}}
\end{figure}

In Figure \ref{dipolefig} we show how the dipole uncertainty 
changes as a function of
survey radius, $R$, smoothed on a scale of $5\hMpc$. 
We have again assumed the linear power spectrum  
suggested by Peacock \& Dodds (1994). At zero radius the uncertainty
tends towards the 1-d rms velocity. This fairly well matches the
observed local group velocity if one assumes that our motion is a 
fair estimate of the 3-d rms velocity. In that case $v_{\rm rms, 1d}=
v_{\rm rms, 3d}/\sqrt{3} \approx 350 \kms$. As the survey radius 
increases the dipole uncertainty decreases. However, in a flux limited 
redshift survey the uncertainty from shot noise increases with radii
(see Section \ref{sncont}), and eventually dominates over the sampling variance.
For the PSCz we find that 
the uncertainties are equal at about $R=150 \hMpc$, thereafter becoming shot-noise
dominated (see Section \ref{sncont}).

For a power--law spectrum, $P(k)=Ak^n$, the integral can be evaluated
and we find
\be
	\lgl v_{\rm cv}^2(0) \rgl = \frac{A}{2^{n+1}\pi} R^{-1-n} 
	\Gamma(n-1) \cos n\pi/2,
\ee
where the spectral slope is in the range $-1<n<1$.

The equations for the variance and correlations can be also
be simplified for power--law spectra. In this 
case the Bessel integral obeys the scaling relations
\ba
	\lefteqn{\int_0^\infty dk k^n j_{\ell-1}^2(kR) =}\nn 
	& & \int_0^\infty dk k^n j_{0}^2(kR) 
	\frac{\Gamma(\ell+(n-3)/2)\Gamma((5-n)/2)}{
	\Gamma(\ell+(3-n)/2)\Gamma((n-1)/2)}.
\ea
Hence the cosmic variance can be evaluated in terms of the dipole uncertainty;
\ba
	\lgl v^2_r(r) \rgl & =& 3 \lgl v^2_r(0) \rgl \sum_{\ell=1}^\infty
	\frac{\ell^2}{2 \ell +1} \left(\frac{r}{R}\right)^{2(\ell-1)} \nn
	& \times & 
	 \frac{\Gamma(\ell+(n-3)/2)\Gamma((5-n)/2)}{
	\Gamma(\ell+(3-n)/2)\Gamma((n-1)/2)}.
\ea
The transverse and total variances can be expressed similarly.
Hence we find that the uncertainty on the velocities in the
survey only scale as the dipole uncertainty for a power--law
spectrum of fluctuations. In addition 
the uncertainty from the error in the dipole can simply be 
subtracted by removing the $\ell=1$ mode for an all-sky redshift survey,
corresponding to moving to the Local Group rest frame. In the 
next section we find that this is generally true, for arbitrary 
power spectra.

\subsection{Transforming to the Local Group rest frame}
\label{dipolesection}

As we shall see it is more accurate to calculate velocities in the 
observer's Local Group rest frame, rather than the CMB frame used so far.
Projecting the local dipole along and arbitrary radial direction 
we find that
\be
	v_r(\0) = \vb(\0).\rbh = \sqrt{\frac{1}{6 \pi^2}} 
	\int_0^\infty\!\! dk k^2 
	\delta_{1 0}(k) k^{-1} j_{0}(kR)
\label{dipole}
\ee
Similarly the transverse dipole projected perpendicular to an 
arbitrary radial vector is 
\be
	v_t^i(\0) =  \sqrt{\frac{1}{6 \pi^2}} \int_0^\infty \!\! dk k^2 
		(\delta_{1 i}(k) - \delta_{1 0}(k) \rbh_i)
		 k^{-1} j_{0}(kR)
\ee
Similar results for the dipole have been found by Webster et al 1997.

With these equations it is straightforward to show that subtracting the 
dipole term is equivalent to ignoring the monopole and dipole terms
in the harmonic summation over $\ell$. This is because the local group
velocity and other velocities are only connected by a dipole term.
In Figure \ref{velext} we plot the uncertainty in the velocity field
in the Local Group rest frame. With the major source of uncertainty, the 
dipole mode, removed we find that the relative uncertainty in the 
velocity field is small compared with the magnitude of the flow fields
we expect.

\subsection{Shot--noise contribution to the reconstructed redshift survey}
\label{sncont}

The shot--noise contribution to the density field can also be calculated
from equation (\ref{eqvel}), taking into account that the autocorrelation
function for Poisson noise is
\be
	\lgl \delta(\r) \delta(\r') \rgl = \frac{1}{\phi(r)} \delta_D(\r-\r'),
\ee
where $\phi(r)$ is the survey selection function. We find that the shot--noise 
variance is (Taylor \& Rowan-Robinson 1993)
\be
	\lgl v_{\rm sn}^2(r) \rgl=\int_0^R \!\! 
	\frac{r^{'2} dr'}{\phi(r')|r^2-r^{'2}|^2} 
\ee
However, this diverges as $r$ approaches $r'$, as infinite variance
is produced by infinity close discrete particles. To avoid this 
problem in the reconstruction one smooths the density field. Applying
a Gaussian filter here we find that the smoothed shot noise variance is
\be
	\lgl v_{\rm sn}^2(r) \rgl = \int_0^R \!\! \frac{ r^{'2} dr'}{\phi(r')}
	\int_{+1}^1 d\mu |\G(\r-\r')|^2 
\ee
where
\be
	\G(\r) = \rbh 
	\Big( \frac{{\rm erf}(r/\sqrt{2}R_s)}{r^2}-\sqrt{\frac{2}{\pi}}
	\frac{e^{-r^2/2R^2_s}}{r R_s}\Big)
\ee
is the Gaussian smoothed Greens function.

The shot noise uncertainty on the velocity can also be decomposed into 
radial and tangential components along the line of sight. The radial term
is 
\be
	\lgl v_{\rm r, sn}^2(r) \rgl = \int_0^R \frac{ r^{'2} dr'}{\phi(r')}
	\int_{+1}^1 d\mu |\rbh.\G(\r-\r')|^2, 	  
\ee
and the tangential component can be found from the relation $
\lgl v_{\rm t, sn}^2 \rgl=\lgl v_{\rm sn}^2 \rgl - \lgl v_{\rm r, sn}^2 \rgl$.

Figure \ref{dipolefig} shows the shot noise contribution to the dipole for the
example of the PSCz. We see that the  shot--noise contribution rises just as the 
sampling variance contribution is falling (Figure \ref{velext}).

\begin{figure}
\vspace{-3.cm}
 \centerline{\epsfig{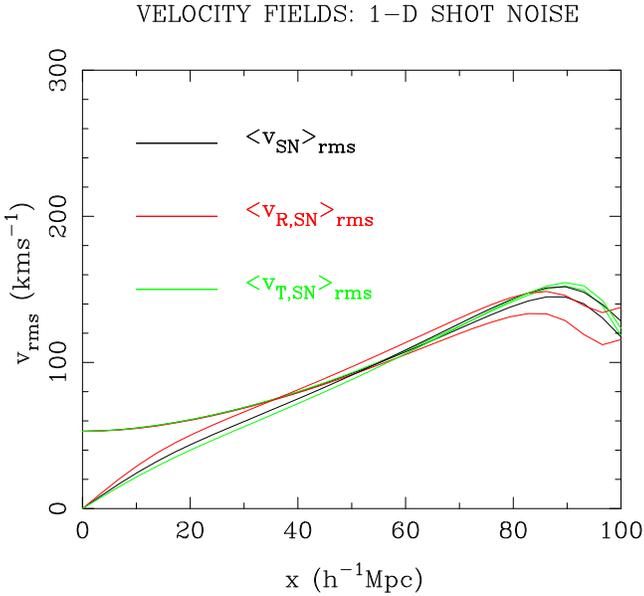}}
\vspace{-3.cm}
 \caption{Shot noise in reconstructed density field due to 
internal galaxy distribution in a redshift survey with radius $R=100h^{-1}$Mpc.}
\end{figure}

Figure 3 shows the radial, transverse and total rms velocity
dispersion due to shot noise in the CMB and Local Group rest frames.
The overall effect of moving to the Local Group rest frame is
marginal, and only significant for velocities at small radii. Hence 
moving to the Local Group rest frame does not affect the shot--noise
contribution to the uncertainties.

\section{Conclusions}
\label{conclusions}

In this paper we have derived the inverse redshift space operator.
By applying this operator to a redshift survey with the correct 
distortion parameter, found either from some other method,
eg a study of the anisotropy of clustering in redshift space, 
or by comparing the reconstructed
velocity field with the true one, a real space map of the density field 
can be recovered. From this one can reconstruct to linear
order the real space potential and velocity fields. We have derived
inverse operators for redshift surveys for observers  in the CMB 
or Local Group rest frames, and we have derived operators which include the 
effects of a selection function.

As a corollary to the calculation of an inverse operator in 
the Local Group rest frame we have also shown how to calculate 
the observer's dipole directly from a redshift survey, without 
having to fully reconstruct the density field, again in either the CMB
or Local group rest frames. Simple tests on an ensemble of Gaussian random
density fields have shown that both the reconstruction of the density 
fields, and the observers dipole are accurate.

To simplify calculations in redshift space we have developed the
formalism using a spherical harmonic representation for the 
fields and the distortion operators. This approach allows one to 
easily estimate the reconstructed field from a spherical harmonic
decomposition of the redshift space density field. We have also found
expressions that relate the real space dipole to
the redshift space density dipole. 

The spherical harmonic representation is also used to estimate the
effects of a finite survey volume on the reconstruction. We find that 
the uncertainty on the dipole mode diverges at the origin, and suggest
that this can be removed by working in the Local Group rest frame.
The quadrupole uncertainty is a constant across the survey volume, while
the sum of the remaining multipoles increases with radius.

The velocity field can also be inferred from a reconstructed 
redshift survey and we have calculated the effects of a finite
survey volume and shot noise on the reconstruction of the velocity 
field, for the special case of no distortion. We have calculated the
sampling variance expected for the velocity field reconstructed 
from a redshift survey and find that the velocity field relative
to the observers motion can be calculated far more accurately than
absolute velocities. Hence one should work in the Local Group 
rest frame when comparing velocities. Calculation of the shot--noise
contribution shows that the shot--noise is fairly insensitive to 
the rest frame.

\bigskip
\noindent{\bf ACKNOWLEDGMENTS}
\bib\strut

\noindent
ANT thanks the PPARC for a research fellowship and HEMV 
thank the PPARC for a studentship.

\bigskip
\noindent{\bf REFERENCES}
\bib \strut

\bib Ballinger W.E., Heavens A.F., Taylor A.N., 1995, MNRAS, 276, L59

\bib Branchini, E., et al, 1998, MNRAS, submitted (astro-ph/9810106)

\bib Croft R.A.C., Gaztanaga E., 1997, MNRAS, 285, 793

\bib Fisher K., et al 1995, MNRAS, 272, 885

\bib Hamilton A.J.S., 1997, in {\em Ringberg Workshop on Large--Scale
Structure}, ed D. Hamilton, Kluwer Academic, astro-ph/9708102

\bib Heavens A.F., Taylor A.N., 1995, MNRAS, 275, 483

\bib Heavens A.F., Taylor A.N., 1997, MNRAS, 290, 456

\bib Kaiser N., et al, 1991, MNRAS, 252, 1

\bib Kaiser N., Lahav O., 1988, in {\em Large--Scale Motions in 
the Universe}, Eds V. Rubin \& G. Coyne, Princeton University Press

\bib Maddox S., et al, 1998, in Proc. MPA/ESO Conf., {\em Evolution
of Large--Scale structure: from Recombination to Garching}, Twin Press

\bib Nusser A., Davis M., 1994, ApJLett, 421, L1

\bib Peacock J.A., Dodds S.J., 1994, MNRAS, 267, 1020

\bib Peebles P.J.E., 1980, {\em The Large--Scale Structure of the Universe},
Princeton University Press, Princeton

\bib Peebles, P.J.E., 1989, ApJLett, 344, L53

\bib Reg\"{o}s E., Szalay A.S., 1989, ApJ, 354, 627

\bib Rowan-Robinson M., et al, 1990, MNRAS, 247, 1

\bib Rowan-Robinson M., et al 1999, MNRAS, to be submitted

\bib Saunders W., et al, 1996, in XXXIIrd Rencontres de Moriond, {\em
	Fundamental Parameters in Cosmology}

\bib Schmoldt I., et al, 1999, MNRAS, in press (astro-ph/9901087)

\bib Shaya E., Peebles P.J.E, Tully R.B. , 1995, ApJ, 454, 15

\bib Szalay A., et al, 1998, Proc. 12th Potsdam Cosmology Workshop 
	{\em Large Scale Structure: Tracks and Traces}, Eds. V. Mueller, 
	S. Gottloeber, J.P. Muecket, J. Wambsganss World Scientific

\bib Tadros H., et al, 1998, MNRAS, in press

\bib Taylor A.N, Hamilton A.J.S., 1996, MNRAS, 282, 767

\bib Taylor A.N., Rowan-Robinson M., 1993, MNRAS, 265, 809

\bib Tegmark M., Bromley B.C., 1995, ApJ, 453, 533

\bib Valentine H.E.M., et al, 1999, MNRAS, to be submitted

\bib Watson G.N., 1966, {\em A Treatise on the Theory of 
Bessel Functions}, Cambridge University Press

\bib Webster M., Lahav O., Fisher K., 1997, MNRAS, 287, 425

\bib Yahil A., et al, 1991, ApJ, 372, 380

\bib Zaroubi S., Hoffman Y., Fisher K., Lahav O., 1995, ApJ, 449, 446

\bib Zaroubi S., Hoffman Y., 1996, ApJ, 462, 25

\end{document}